\documentclass[12pt, a4paper]{article}
\usepackage{amsfonts}
\usepackage{latexsym}
\usepackage{amssymb}
\usepackage{mathrsfs}
\usepackage{amsmath}
\usepackage{graphicx}
\usepackage{cite}

\begin{document}

\begin{flushleft}
{\large
\textbf{Multiscale Analysis of Information Dynamics for Linear Multivariate Processes}
}
\\
\vspace{5pt}
{\small
Luca Faes$^{1,\ast}$,
Alessandro Montalto$^{2}$, 
Sebastiano Stramaglia$^{3}$,
Giandomenico Nollo$^{1}$,
Daniele Marinazzo$^{2}$
}
\\
\vspace{5pt}
{\scriptsize
\bf{1} BIOtech, Dept. of Industrial Engineering, University of Trento, and IRCS-PAT FBK, Trento, Italy
\\
\bf{2} Data Analysis Department, Ghent University, Ghent, Belgium
\\
\bf{3} Dipartimento Interateneo di Fisica, University of Bari, and INFN Sezione di Bari, Italy
\\
$\ast$ E-mail: faes.luca@gmail.com
}
\end{flushleft}

\thispagestyle{empty}
\pagestyle{empty}

\begin{abstract}

In the study of complex physical and physiological systems represented by multivariate time series, an issue of great interest is the description of the system dynamics over a range of different temporal scales. While information-theoretic approaches to the multiscale analysis of complex dynamics are being increasingly used, the theoretical properties of the applied measures are poorly understood. This study introduces for the first time a framework for the analytical computation of information dynamics for linear multivariate stochastic processes explored at different time scales. After showing that the multiscale processing of a vector autoregressive (VAR) process introduces a moving average (MA) component, we describe how to represent the resulting VARMA process using state-space (SS) models and how to exploit the SS model parameters to compute analytical measures of information storage and information transfer for the original and rescaled processes. The framework is then used to quantify multiscale information dynamics for simulated unidirectionally and bidirectionally coupled VAR processes, showing that rescaling may lead to insightful patterns of information storage and transfer but also to potentially misleading behaviors.

\end{abstract}

\section{Introduction}

Several physiological systems, including the brain and the cardiovascular system, coordinate their activity according to regulatory mechanisms operating across multiple temporal scales
\cite{Ivanov1999461, Thakor2009}. Due to this multiscale behavior, the output signals of these systems (e.g., the EEG or cardiovascular variability time series) need to be analyzed through scaling techniques to get full insight about the system dynamics. A typical approach is to resample the originally measured physiological time series at various temporal scales, yielding a collection of rescaled series from which various dynamical measures can be calculated. Exploiting information-theoretic functionals that may be subsumed within the framework of information dynamics \cite{Faes2015277}, this approach has been followed both to describe the individual dynamics of single time series through the so-called multiscale entropy \cite{costa2002multiscale}, and to explore the joint dynamics of multiple time series through the multiscale transfer entropy (TE) \cite{Lungarella2007}.

In spite of its potential, the computation of multiscale measures of information dynamics may be complicated by theoretical and practical issues \cite{Barnett2011404} \cite{Valencia20092202}. These issues arise from the procedure for the generation of the rescaled time series, which essentially consists in a filtering step eliminating the fast temporal scales (usually performed through averaging) followed by a downsampling step coarse-graining the time series around the selected scale. While it is expected that these two steps may be problematic, their impact on the computation of multiscale information dynamics has never been investigated systematically. To fill this gap, the present study introduces a framework for the analytical computation of information dynamics for linear Gaussian dynamic processes subjected to averaging and downsampling. The framework is based on the theory of state-space (SS) models, and builds on very recent theoretical results \cite{barnett2015granger} \cite{solo2015state} to study the exact values of information storage (storage entropy, SE) and information transfer (TE) for coupled processes observed at different time scales. While this study concentrates on the theoretical formulation and analysis of simulated linear processes, future extensions will be devoted to practical estimation, study of nonlinear dynamics and  application to real time series.

\section{Multiscale Representation of Linear Processes} \label{sec:MS}
Let us consider a set of \textit{M} time series of length \textit{N},
$y_{m,n}, m=1,\ldots,M; n=1,\ldots,N$, as a finite length realization of the zero mean stationary vector stochastic process $Y_n=[y_{1,n}\cdots y_{M,n}]^T$. In the linear signal processing framework, the process is classically described as a Vector Autoregressive (VAR) process of order $p$:

\begin{equation} \label{eq:VAR}
Y_n = \sum_{k=1}^{p}{\mathbf{A}_k Y_{n-k} + U_n}
\end{equation}
where $A_k$ are $M\times M$  matrices of coefficients, and
$U_n=[u_{1,n}\cdots u_{M,n}]^T$ is a vector of $M$ zero mean Gaussian processes with
covariance matrix {\boldmath$\Sigma$}$\equiv$$\mathbb{E}[U_nU_n^T]$.

According to the traditional procedure for multiscale analysis \cite{costa2002multiscale}, each scalar process $y_m$ can be rescaled using an integer scale factor $\tau$ to get the process $\bar{y}_m$:

\begin{equation} \label{eq:MSY}
\bar{y}_{m,n} = \frac{1}{\tau} \sum_{l=0}^{\tau-1}{y_{m,n\tau-l}} , n=1,\ldots,N/\tau
\end{equation}

The change of scale in (\ref{eq:MSY}) corresponds to transform the original process $Y$ through a two step procedure that consists of the following \textit{averaging} and \textit{downsampling} steps, yielding respectively the processes $\tilde{Y}$ and $\bar{Y}$:

\begin{subequations} \label{eq:AvgDws}
\begin{align}
		\tilde{Y}_n &= \frac{1}{\tau} \sum_{l=0}^{\tau-1}{Y_{n-l}} , n=\tau,\ldots,N  \label{eq:Avg} \\
		\bar{Y}_n &= \tilde{Y}_{n\tau} , n=1,\ldots,N/\tau \label{eq:Dws}
\end{align}
\end{subequations}

Now, substituting (\ref{eq:VAR}) in (\ref{eq:Avg}), one can show that the averaging step yields the following process representation:

\begin{equation} \label{eq:VARMAAvg}
\tilde{Y}_n = \sum_{k=1}^{p}{\mathbf{A}_k \tilde{Y}_{n-k}} + \sum_{l=0}^{\tau-1}{\mathbf{B}_l U_{n-l}}
\end{equation}

where $\mathbf{B}_l=1 / \tau I_M$ for each $l=0,\ldots , \tau -1$ ($I_M$ is the $M\times M$ identity matrix). This shows that the change of scale introduces a moving average (MA) component of order $q=\tau-1$ in the original VAR$(p)$ process, transforming it into a VARMA$(p,q)$ process. As we will show in the next Section, the downsampling step (\ref{eq:Dws}) keeps the VARMA representation altering the model parameters.

\section{State Space Models} \label{sec:SS}

\subsection{Formulation of SS Models} \label{sec:SSformul}
The general linear state space (SS) model describing an observed vector process $Y$ is in form:
\begin{subequations} \label{eq:eqSS}
\begin{align}
		X_{n+1} &= \mathbf{A} X_{n} + W_{n} \label{eq:eqSSstate} \\
		Y_n &= \mathbf{C} X_{n} + V_{n} \label{eq:eqSSobs}
\end{align}
\end{subequations}

where the state equation (\ref{eq:eqSSstate}) describes the update of the $L-$dimensional state (unobserved) process through the $L \times L$ matrix $\mathbf{A}$, and the observation equation (\ref{eq:eqSSobs})
describes the instantaneous mapping from the state to the observed process through the $M \times L$ matrix $\mathbf{C}$. $W_n$ and $V_n$ are zero-mean white noise processes with covariances
{\boldmath$\Xi$}$\equiv$$\mathbb{E}[W_nW_n^T]$ and {\boldmath$\Psi$}$\equiv$$\mathbb{E}[V_nV_n^T]$, and cross-covariance {\boldmath$\Upsilon$}$\equiv$$\mathbb{E}[W_nV_n^T]$. Thus, the parameters of the SS model (\ref{eq:eqSS}) are ($\mathbf{A},\mathbf{C},\mathbf{\Xi},\mathbf{\Psi},\mathbf{\Upsilon}$).

Another possible SS representation is that evidencing the \textit{innovations} 
$E_n=Y_n-\mathbb{E}[Y_n|Y_n^-]$, i.e. the residuals of the linear regression of $Y_n$ on its infinite past
$Y_n^- = [Y_{n-1}^T Y_{n-2}^T \cdots]^T$. This new SS representation, usually referred to as innovations form SS model (ISS), is characterized by the state process
$Z_n=\mathbb{E}[X_n|Y_n^-]$ and by the $L \times M$ Kalman Gain matrix $\mathbf{K}$:
\begin{subequations} \label{eq:eqISS}
\begin{align}
		Z_{n+1} &= \mathbf{A} Z_{n} + \mathbf{K} E_{n} \label{eq:eqISSstate} \\
		Y_n &= \mathbf{C} Z_{n} + E_{n} \label{eq:eqISSobs}
\end{align}
\end{subequations}

The parameters of the ISS model (\ref{eq:eqISS}) are ($\mathbf{A},\mathbf{C},\mathbf{K},\mathbf{\Phi}$), 
where $\mathbf{\Phi}$ is the covariance of the innovations, {\boldmath$\Phi$}$\equiv$$\mathbb{E}[E_nE_n^T]$.
Note that the ISS (\ref{eq:eqISS}) is a special case of (\ref{eq:eqSS}) in which $W_n=\mathbf{K} E_n$ and
$V_n=E_n$, so that $\mathbf{\Xi}=\mathbf{K}\mathbf{\Phi}\mathbf{K}^T$, $\mathbf{\Psi}=\mathbf{\Phi}$
and $\mathbf{\Upsilon}=\mathbf{K}\mathbf{\Phi}$.

Given an SS model in the form (\ref{eq:eqSS}), the corresponding ISS model (\ref{eq:eqISS}) can be identified
by solving a so-called discrete algebraic Ricatti equation (\textit{DARE}) formulated in terms of the state error variance matrix $\mathbf{P}$:
\begin{equation} \label{eq:DARE}
\begin{aligned}
	\mathbf{P} &= \mathbf{A}\mathbf{P}\mathbf{A}^T + \mathbf{\Xi} \\
						&-(\mathbf{A}\mathbf{P}\mathbf{C}^T+\mathbf{\Upsilon})
				(\mathbf{C}\mathbf{P}\mathbf{C}^T+\mathbf{\Psi})^{-1} (\mathbf{C}\mathbf{P}\mathbf{A}^T+\mathbf{\Upsilon}^T)
\end{aligned}
\end{equation}
Under some assumptions \cite{solo2015state}, the \textit{DARE} (\ref{eq:DARE}) has an unique stabilizing solution, from which the Kalman gain and innovation covariance can be computed as
\begin{equation} \label{eq:SStoISS}
\begin{aligned}
						\mathbf{\Phi} &= \mathbf{C}\mathbf{P}\mathbf{C}^T + \mathbf{\Psi} \\
						\mathbf{K} &= (\mathbf{A}\mathbf{P}\mathbf{C}^T + \mathbf{\Upsilon})\mathbf{\Phi}^{-1}
\end{aligned}
\end{equation}

\subsection{SS Models for Averaged and Downsampled Processes} \label{sec:SSAvgDws}
Exploiting the close relation between VARMA models and SS models, first we show how to convert the VARMA model
(\ref{eq:VARMAAvg}) into an ISS model in the form of (\ref{eq:eqISS}) that describes the 
averaged process $\tilde{Y_n}$. To do this, we exploit the Aoki's method \cite{Aoki1991} defining the state process $\tilde{Z}_n=[Y_{n-1}^T \cdots Y_{n-p}^T U_{n-1}^T \cdots U_{n-q}^T]^T$ that, together with $\tilde{Y_n}$, obeys the state equations
(\ref{eq:eqISS}) with parameters ($\tilde{\mathbf{A}},\tilde{\mathbf{C}},
\tilde{\mathbf{K}},\tilde{\mathbf{\Phi}}$), where

\[\tilde{\mathbf{A}}
=
\begin{bmatrix}
    \mathbf{A}_1&\cdots&\mathbf{A}_{p-1}&\mathbf{A}_p & \mathbf{B}_1&\cdots&\mathbf{B}_{q-1}&\mathbf{B}_q \\
		\mathbf{I}_M&\cdots&\mathbf{0}_M    &\mathbf{0}_M & \mathbf{0}_M&\cdots&\mathbf{0}_M    &\mathbf{0}_M \\
		\vdots      &      &\vdots          &\vdots       & \vdots      &      &\vdots          &\vdots		    \\
		\mathbf{0}_M&\cdots&\mathbf{I}_M    &\mathbf{0}_M & \mathbf{0}_M&\cdots&\mathbf{0}_M    &\mathbf{0}_M \\
		\mathbf{0}_M&\cdots&\mathbf{0}_M    &\mathbf{0}_M & \mathbf{0}_M&\cdots&\mathbf{0}_M    &\mathbf{0}_M \\
		\mathbf{0}_M&\cdots&\mathbf{0}_M    &\mathbf{0}_M & \mathbf{I}_M&\cdots&\mathbf{0}_M    &\mathbf{0}_M \\
		\vdots      &      &\vdots          &\vdots       & \vdots      &      &\vdots          &\vdots		    \\
		\mathbf{0}_M&\cdots&\mathbf{0}_M    &\mathbf{0}_M & \mathbf{0}_M&\cdots&\mathbf{I}_M    &\mathbf{0}_M
\end{bmatrix}
\]
\[\tilde{\mathbf{C}}
=
\begin{bmatrix}
	\mathbf{A}_1&\cdots&\mathbf{A}_p & \mathbf{B}_1&\cdots&\mathbf{B}_{q}
\end{bmatrix}
\]
\[\tilde{\mathbf{K}}
=
\begin{bmatrix}
	\mathbf{I}_M & \mathbf{0}_{M\times M(p-1)} &\mathbf{B}_0^{-T} & \mathbf{0}_{M\times M(q-1)}
\end{bmatrix}^T
\]

and $\tilde{\mathbf{\Phi}} = \mathbf{B}_0 \mathbf{\Sigma} \mathbf{B}_0^T$, where $\tilde{\mathbf{\Phi}}$
is the covariance of the innovations $\tilde{E}_n=\mathbf{B}_0 U_n$.

Now we turn to show how the downsampled process $\bar{Y}_n$ can be represented through an ISS model directly from the ISS formulation of the averaged process $\tilde{Y}_n$. According to a very recent result (theorem III in \cite{solo2015state}), we have that the process $\bar{Y}_n=\tilde{Y}_{n\tau}$ has an ISS representation with state process
$\bar{Z}_n=\tilde{Z}_{n\tau}$, innovation process $\bar{E}_n=\tilde{E}_{n\tau}$, and parameters
($\bar{\mathbf{A}},\bar{\mathbf{C}},\bar{\mathbf{K}},\bar{\mathbf{\Phi}}$), where 
$\bar{\mathbf{A}}=\tilde{\mathbf{A}}^\tau$, $\bar{\mathbf{C}}=\tilde{\mathbf{C}}$, and where 
$\bar{\mathbf{K}}$ and $\bar{\mathbf{\Phi}}$ are obtained solving the \textit{DARE} (\ref{eq:DARE},\ref{eq:SStoISS}) for the SS model ($\bar{\mathbf{A}},\bar{\mathbf{C}},\mathbf{\Xi}_\tau,
\tilde{\mathbf{\Phi}},\mathbf{\Upsilon}_\tau$) with
\begin{equation} \label{eq:QSdown}
\begin{aligned}
	\mathbf{\Upsilon}_\tau &= \tilde{\mathbf{A}}^{\tau-1} \tilde{\mathbf{K}}\tilde{\mathbf{\Phi}} \\
	\mathbf{\Xi}_\tau &= \tilde{\mathbf{A}} \mathbf{\Xi}_{\tau-1} \tilde{\mathbf{A}}^T
	+ \tilde{\mathbf{K}}\tilde{\mathbf{\Phi}}\tilde{\mathbf{K}}^T, \tau\geq 2 \\
	\mathbf{\Xi}_1 &= \tilde{\mathbf{K}} \tilde{\mathbf{\Phi}} \tilde{\mathbf{K}}^T, \tau=1
\end{aligned}
\end{equation}

\begin{figure}
\centering
\includegraphics[width=10cm]{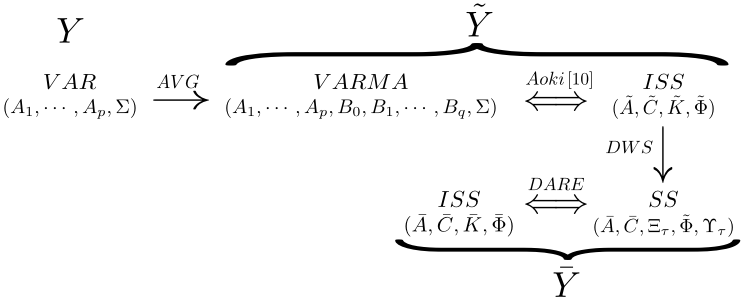}
\caption{Schematic representation of the parametric representation of linear multivariate processes. See text for details.}
\label{fig:scheme}
\end{figure}

\section{Multiscale Information Dynamics} \label{sec:ID}
Fig. \ref{fig:scheme} depicts the relations and parametric representations of the original process $Y$, the averaged process $\tilde{Y}$, and the downsampled process $\bar{Y}$.
As seen up to now, the averaging (AVG) over segments of length $\tau$ applied to a VAR($p$) process yields a VARMA($p, \tau-1$) process, which is equivalent to an ISS process \cite{Aoki1991}, and the subsequent downsampling (DWS) yields a different SS process, which in turn can be converted to the ISS form using solving the \textit{DARE} (Fig. 1). Thus, both the averaged process $\tilde{Y}_n$ and the downsampled process $\bar{Y}_n$ can be represented as ISS processes with parameters 
($\tilde{\mathbf{A}},\tilde{\mathbf{C}},\tilde{\mathbf{K}},\tilde{\mathbf{\Phi}}$) 
and ($\bar{\mathbf{A}},\bar{\mathbf{C}},\bar{\mathbf{K}},\bar{\mathbf{\Phi}}$)
which can be derived analytically from the knowledge of the parameters ($\mathbf{A}_1,\ldots, \mathbf{A}_p, \Sigma$) of the original process and of the scale factor $\tau$. In this section we show how to compute analytically the measures of information dynamics starting from the ISS model parameters, thus opening the way to the analytical computation of these measures for multiscale (averaged and downsampled) processes. 

Given a generic vector observation process $Y_n$, let us consider the scalar subprocess $y_{j,n}$ as the
\textit{target}, and the $(M-1)-$dimensional vector $Y_{i,n}=Y_n\backslash y_{j,n}$ as the \textit{driver}
($i=\{1\cdots,M \} \backslash j$). In the framework of information dynamics \cite{Faes2015277}, the predictive information of the target of a multivariate process, $P_j$, measures how much of the information carried by $y_{j,n}$ can be predicted from the knowledge of $Y_n^-$. This amount can be decomposed as the sum of the information storage $S_j$ and the information transfer $T_{i \rightarrow j}$, quantifying respectively the amount of information carried by $y_{j,n}$ that can be predicted from its own past $y_{j,n}^-$ and the additional amount that can be predicted from the whole past $Y_n^-$. The information storage and transfer are quantified by the so-called storage entropy (SE) and transfer entropy (TE) \cite{Faes2015} which, for linear Gaussian processes, are given by:
\begin{subequations} \label{eq:SETE}
\begin{align}
		S_j=\frac{1}{2} ln \frac{\lambda_j}{\lambda_{j|j}} \label{eq:SE} \\
		T_{i \rightarrow j}=\frac{1}{2} ln \frac{\lambda_{j|j}}{\lambda_{j|ij}} \label{eq:TE}
\end{align}
\end{subequations}
where $\lambda_j=\mathbb{E}[y_{j,n}^2]$ is the variance of the target process, and 
$\lambda_{j|j}=\mathbb{E}[e_{j|j,n}^2]$ and $\lambda_{j|ij}=\mathbb{E}[e_{j|ij,n}^2]$ are the partial variance of the target given its own past, $e_{j|j,n}=y_{j,n}-\mathbb{E}[y_{j,n}|y_{j,n}^-]$, and the partial variance of the target given the past of the whole process, $e_{j|ij,n}=y_{j,n}-\mathbb{E}[y_{j,n}|Y_n^-]$.

Now we report how to compute the variances appearing in (\ref{eq:SETE}) from the parameters of an ISS model in the form of (\ref{eq:eqISS}). First, we note that the variance of $e_{j|ij,n}$ is simply the $j-th$ diagonal element of the innovation covariance: $\lambda_{j|ij}=\mathbf{\Phi}(j,j)$.
The variance of $y_{j,n}$ corresponds to the $j-th$ diagonal element of the zero-lag autocovariance of the whole process $\mathbf{\Gamma} \equiv \mathbb{E}[Y_n Y_n^T]$: $\lambda_{j}=\mathbf{\Gamma}(j,j)$;
for an ISS process, the latter can be computed as
$\mathbf{\Gamma}=\mathbf{C}\mathbf{\Omega}\mathbf{C}^T+\mathbf{\Phi}$, where $\mathbf{\Omega}=\mathbb{E}[Z_nZ_n^T]$ satisfies the discrete Lyapunov equation
$\mathbf{\Omega}=\mathbf{A}\mathbf{\Omega}\mathbf{A}^T+\mathbf{K}\mathbf{\Phi}\mathbf{K}^T$.
Computation of the partial variance of the target given its past is less straightforward, involving the formation of a subprocess of the original ISS process. Specifically, one needs to consider the submodel with state equation (\ref{eq:eqISSstate}) and observation equation 
\begin{equation} \label{eq:SSreduced}
		y_{j,n}=\mathbf{C}^{(j)} Z_n + e_{j,n}
\end{equation}
where $\mathbf{C}^{(j)}$ is the $j-th$ row of $\mathbf{C}$. The submodel (\ref{eq:eqISSstate}, \ref{eq:SSreduced}) is \textit{not} in innovations form, but is rather an SS model with parameters ($\mathbf{A},\mathbf{C}^{(j)},\mathbf{K}\mathbf{\Phi}\mathbf{K}^T,
\mathbf{\Phi}(j,j),\mathbf{K}\mathbf{\Phi}^{T^{(j)}}$). As such, solving the \textit{DARE} (\ref{eq:DARE},\ref{eq:SStoISS}) it can be converted to an ISS model with innovation covariance $\lambda_{j|j}$.

\begin{figure}
\centering
\includegraphics[width=10cm]{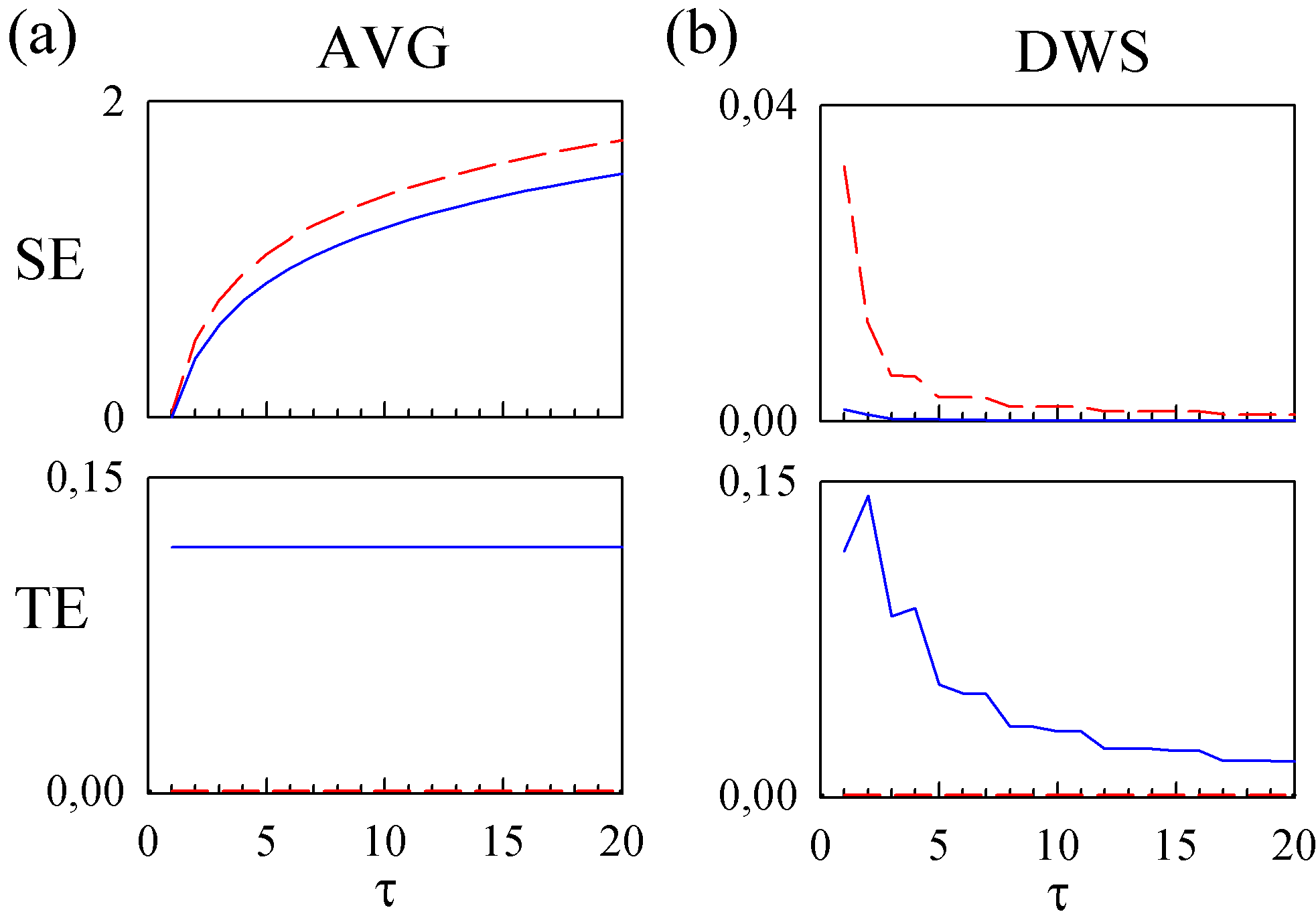}
\caption{Multiscale information dynamics for the unidirectionally coupled VAR process (\ref{eq:simuVAR}). Plots depict the information storage (SE) and transfer (TE) computed after averaging (AVG) and downsampling (DWS) the process at scale $\tau$; red-dashed: $S_1, T_{2 \rightarrow 1}$, blue-solid: $S_2, T_{1 \rightarrow 2}$. (a,b) parameter $a_1=0.25$; (c,d) parameter $a_1=0.95$.}
\label{fig:simu1}
\end{figure}
\begin{figure}
\centering
\includegraphics[width=10cm]{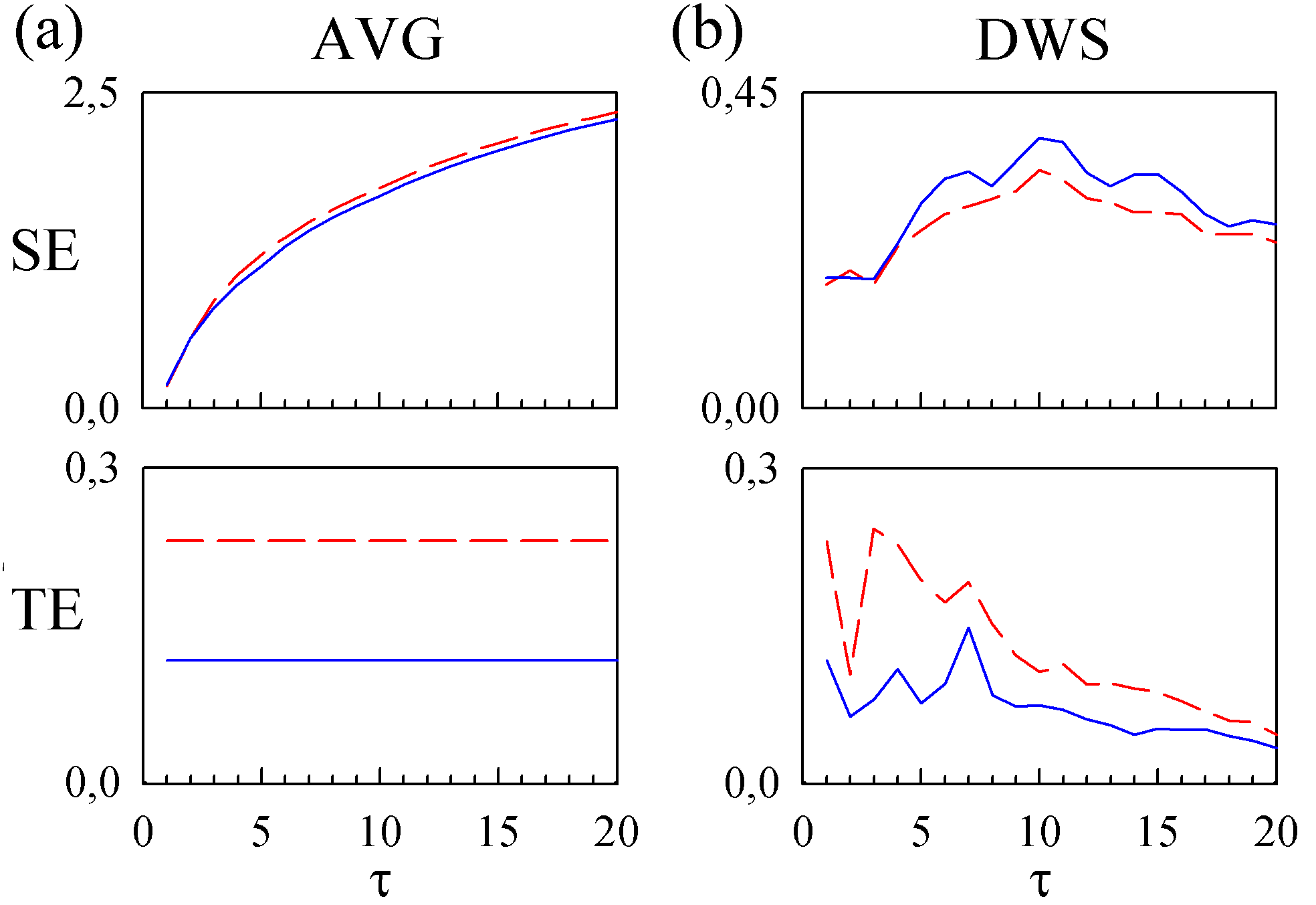}
\caption{Multiscale information dynamics for the bidirectionally coupled VAR process (\ref{eq:simuVAR}). Plots and symbols are as in Fig. \ref{fig:simu1}.}
\label{fig:simu2}
\end{figure}

\section{Simulation Experiment} \label{sec:simu}
In order to study the multiscale patterns of information dynamics for linear interacting processes, we analyze the bivariate VAR process with equations:
\begin{subequations} \label{eq:simuVAR}
\begin{align}
		y_{1,n} &= a_1 y_{1,n-b_1} + c_1 y_{2,n-d_1} + u_{1,n} \\
		y_{2,n} &= a_2 y_{2,n-b_2} + c_2 y_{1,n-d_2} + u_{2,n}
\end{align}
\end{subequations}
with iid noise processes $u_{1,n}, u_{2,n} \sim \mathcal{N}(0,1)$ so that $\mathbf{\Sigma}=\mathbf{I}_2$. The parameters in (\ref{eq:simuVAR}) are set to generate autonomous dynamics with strength $a_i$ and lag $b_i$ for each scalar process $y_i$, and causal interactions with strength $c_i$ and lag $d_i$ from $y_j$ to $y_i$
($i,j=1,2$). We consider two parameter configurations: unidirectional interaction $y_1 \rightarrow y_2$, obtained setting $c_1=0$ and $c_2=0.5, d_2=2$, where also autonomous dynamics were generated for $y_1$ ($a_1=0.25, b_1=1$) but not for $y_2$ ($a_2=0$); bidirectional interactions between processes with autonomous dynamics ($a_1=0.25, b_1=2; a_2=0.25, b_2=5$) obtained setting $c_2=0.5, d_2=7$ (direction $y_1 \rightarrow y_2$) and $c_1=0.75, d_1=3$ (direction $y_2 \rightarrow y_1$).


The results of multiscale analysis of SE and TE performed for the two configurations are shown in Figs. \ref{fig:simu1}, \ref{fig:simu2}. The values of information dynamics for the original processes, reported in the figures for $\tau=1$, indicate that the SE reflects auto-dependencies in the target process (e.g., $S_1>S_2$ in Fig. \ref{fig:simu1} where $a_1>a_2$, and $S_1=S_2$ in Fig. \ref{fig:simu2} where $a_1=a_2$), and that the TE reflects causal interactions from driver to target (e.g., $T_{2 \rightarrow 1}=0$ in Fig. \ref{fig:simu1} where $c_1=0$ and  $T_{2 \rightarrow 1}>T_{1 \rightarrow 2}$ in Fig. \ref{fig:simu2} where $c_1>c_2$). The averaging procedure associated with the change of scale always leads to a progressive increase of the information stored in each individual process (Figs. \ref{fig:simu1}a, \ref{fig:simu2}a). Moreover, averaging does not alter the amount of information transferred between the processes, as documented by the constant values of the TE across scales observed in all configurations. The downsampling step introduces more substantial alterations in the patterns of information dynamics. The information storage is reduced substantially and reflects the multiscale regularity of each individual process, with higher SE around the scales at which the processes exhibit their lagged interactions (i.e., very low lags in Fig.\ref{fig:simu1}(b) and higher lags in Fig. \ref{fig:simu2}(b)). The information transfer reflects causal interactions between the processes at different time scales, with the TE showing a peak at the lags of the imposed causal interactions (i.e., $\tau=2$ for $T_{1 \rightarrow 2}$ in Fig. \ref{fig:simu1}(b), $\tau=7$ for $T_{1 \rightarrow 2}$ and $\tau=3$ for $T_{2 \rightarrow 1}$ in Fig. \ref{fig:simu2}(b)).

\begin{figure}
\centering
\includegraphics[width=10cm]{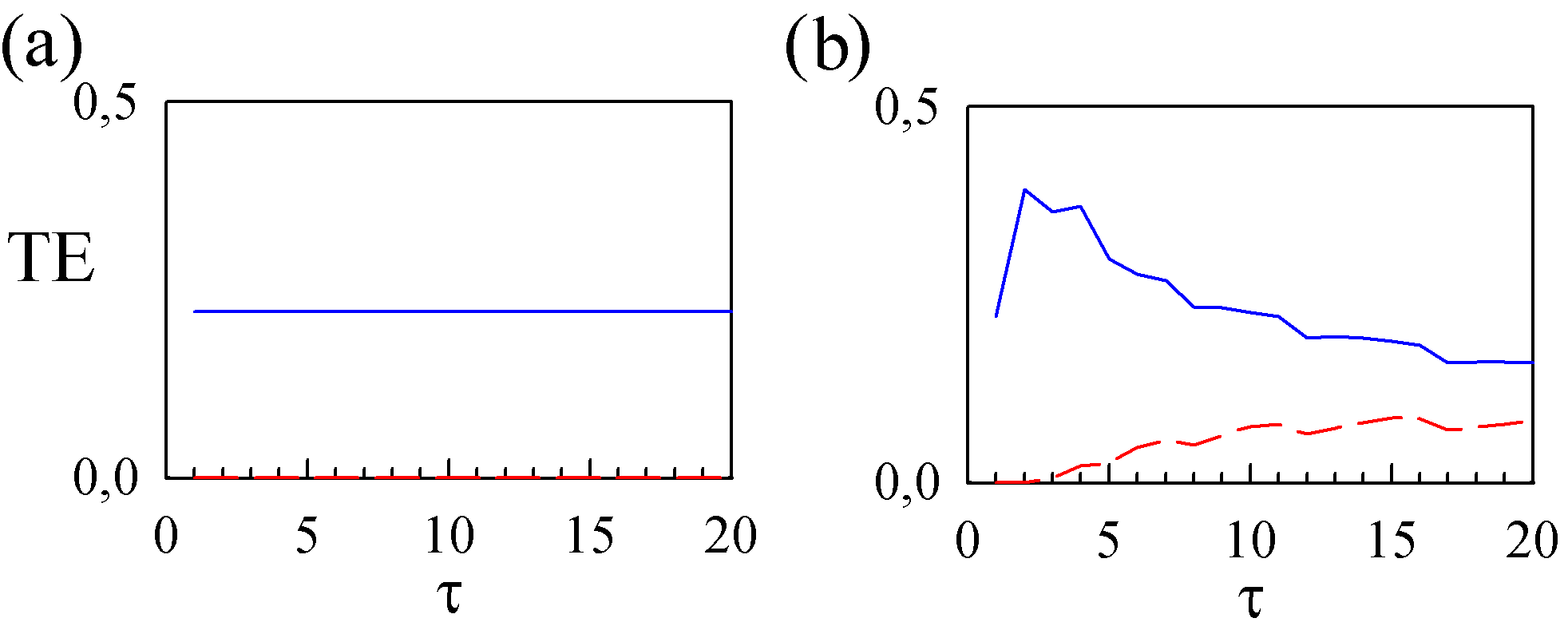}
\caption{Multiscale information dynamics for the unidirectionally coupled VAR process (\ref{eq:simuVAR}) with stronger driver autonomous dynamics. Plots and symbols are as in Fig. \ref{fig:simu1}.}
\label{fig:simu3}
\end{figure}

The behaviors described above are general, in the sense that they were observed also for different parameter configurations. Nevertheless, some particular parameter settings led to unexpected, potentially misleading results. An example is reported in Fig. \ref{fig:simu3}, showing the information transfer  computed for the first configuration (unidirectional coupling) but with stronger autonomous dynamics of $y_1$ ($a_1=0.95)$. In this case the TE $T_{1 \rightarrow 2}$ still shows a peak at the scale corresponding to the lag of the imposed causal relation ($\tau=d_2=2$), but a significant TE emerges at large scales along the uncoupled direction
($T_{2 \rightarrow 1}>0$ for $\tau > 2$).

\section{Conclusions} \label{sec:conclusion}
We presented a framework for the multiscale computation of the information stored and transferred in multivariate linear processes, assessed respectively through the SE and TE measures, starting from the parameters of the VAR model describing the process and from the scale factor $\tau$.

Our simulation results show that the first step of multiscale analysis, i.e. the averaging of each individual process across $\tau$ consecutive points, introduces an auto-correlation in the process that is reflected by the progressive increase with $\tau$ of the SE. Moreover, as this step leaves the coefficients regulating the linear interaction across the processes unchanged, the TE does not vary with $\tau$; this result is related to the invariance of Granger causality with filtering \cite{Barnett2011404}.

The second analysis step, i.e. the downsampling of the averaged process at fixed time intervals $\tau$, removes the autocorrelation of the innovations inflating the SE, thus allowing a more informative evaluation of the multiscale complexity of the individual time series \cite{costa2002multiscale}. Moreover, this step makes the TE scale-dependent, with a peak shown at the time scale corresponding to the lags of the causal interactions occurring between the processes. A negative behavior is the possible occurrence of spurious TE at scales much higher than the true coupling delays.

These results suggest that the multiscale analysis of information storage and information transfer can be useful to shed light on patterns of regularity and causality of coupled dynamic processes which are not fully disclosed working at one single time scale, but can also provide patterns with difficult physical interpretation.

\section*{Acknowledgments}
Research supported by Healthcare Research and Innovation Program, IRCS-PAT-FBK, Trento.

\addtolength{\textheight}{-12cm}   









\bibliographystyle{IEEEtran}
\bibliography{FaesEMBC2016}

\end{document}